\documentclass[11pt,twoside]{article}

\usepackage{prcsa2008}
\usepackage{epsf}
\usepackage{lscape}
\usepackage{graphicx}

\begin{document}
\title{Jets and environment of microquasars }   
\author{S. N. Zhang\altaffilmark{1}; J. F. Hao\altaffilmark{1}}   
\altaffiltext{1}{Department of Physics and Tsinghua Center for
Astrophysics, Tsinghua University, Beijing 100084, China. zhangsn@tsinghua.edu.cn, jingfang.hao@hotmail.com}    

\begin{abstract} 
Two relativistic X-ray jets have been detected with the \textit{Chandra} X-ray
observatory from the black hole X-ray transient XTE J1550-564. We report a full
analysis of the evolution of the two jets with a gamma-ray burst external shock model.
A plausible scenario suggests a cavity outside the central source and the jets first
travelled with constant velocity and then are slowed down by the interactions between
the jets and the interstellar medium (ISM). The best fitted radius of the cavity is
$\sim$0.36 pc on the eastern side and $\sim$0.46 pc on the western side, and the
densities also show asymmetry, of $\sim$0.015 cm$^{-3}$ on the east to $\sim$0.21
cm$^{-3}$ on the west. A large scale low density region is also found in another
microquasar system, H 1743-322. These results are consistent with previous suggestions
that the environment of microquasars should be rather vacuous, compared to the normal
Galactic environment. A generic scenario for microquasar jets is proposed, classifying
the observed jets into three main categories, with different jet morphologies (and
sizes) corresponding to different scales of vacuous environments surrounding them.
\end{abstract}


\section{INTRODUCTION}
Microquasars are well known miniatures of quasars, with a central black hole (BH), an
accretion disk and two relativistic jets very similar to those found in the centers of
active galaxies, only on much smaller scales (Mirabel $\&$ Radr\'{\i}guez 1999). Since
discovered in 1992, radio jets have been observed in several BH binary systems and some
of them showed apparent superluminal features. In the two well known microqusars, GRS
1915+105 (Mirabel $\&$ Radr\'{\i}guez 1999) and GRO J1655-40 (Tingay et al.1995;
Hjellming $\&$ Rupen 1995), relativistic jets with actual velocities greater than
0.9$c$ were observed. In some other systems, small-size ``compact jets", e.g. Cyg X-1
(Stirling et al. 2001), and large scale diffuse emission, e.g. SS433 (Dubner et al.
1998), were also detected.

XTE J1550-564 was discovered with RXTE during its strong X-ray outburst on September 7,
1998 (Smith 1998). It is believed to be an X-ray binary system at a distance of
$\sim$5.3 kpc, containing a black hole of 10.5$\pm$1.0 solar masses and a low mass
companion star (Orosz et al. 2002). Soon after the discovery of the source, a jet
ejection with an apparent velocity greater than 2$c$ was reported (Hannikainen et al.
2001). In the period between 1998 and 2002, several other outbursts occurred but no
similar radio and X-ray flares were detected again in these outbursts (Tomsick et al.
2003).

With the help of the \textit{Chandra} satellite, Corbel et al (2002) found two large
scale X-ray jets lying to the east and the west of the central source, which were also
in good alignment with the central source. The eastern jet has been detected first in
2000 at a projected distance of $\sim$21$\arcsec$ from the central black hole. Two
years later, the jet could only be seen marginally in the X-ray image, while a western
counterpart became visible at $\sim$22$\arcsec$ on the other side. The corresponding
radio maps are consistent with the X-ray observations (Corbel et al. 2002).

There are altogether eight two-dimentional imaging observations of XTE J1550-564 in
\textit{Chandra} archive during June 2000 and October 2003 (henceforth observations
1$\sim$8). Here we report a full analysis of these X-ray data, together with the
kinematic and spectral evolution fittings for all these observations.

\section{OBSERVATIONS of XTE J1550-564}

The basic information of observations 1$\sim$8 is listed in Table 1, including the
observation ID, date, and the angular separation between the eastern and western jets
and the central source. The positions are obtained by the \textit{Chandra} Interactive
Analysis of Observations (CIAO) routine $wavdetect$ (Freeman et al. 2002). In
observations 5 and 6, no X-ray source is detected by $wavdetect$ at the position of the
eastern jet. However, from the smoothed images (Fig.1), a weak source could be
recognized in observation 6. We thus select the center of the strongest emission region
as one data point. We calculate the source centroid for the central source and the
X-ray jet respectively and for all the five observations, the calculated position
changed by less than 0.5\arcsec. Therefore, an upper limit of 0.5$\arcsec$ is set for
the error of the jet distance.

\begin{figure}
\begin{minipage}[b]{.5\textwidth}
    \centering
    \includegraphics[scale=0.3]{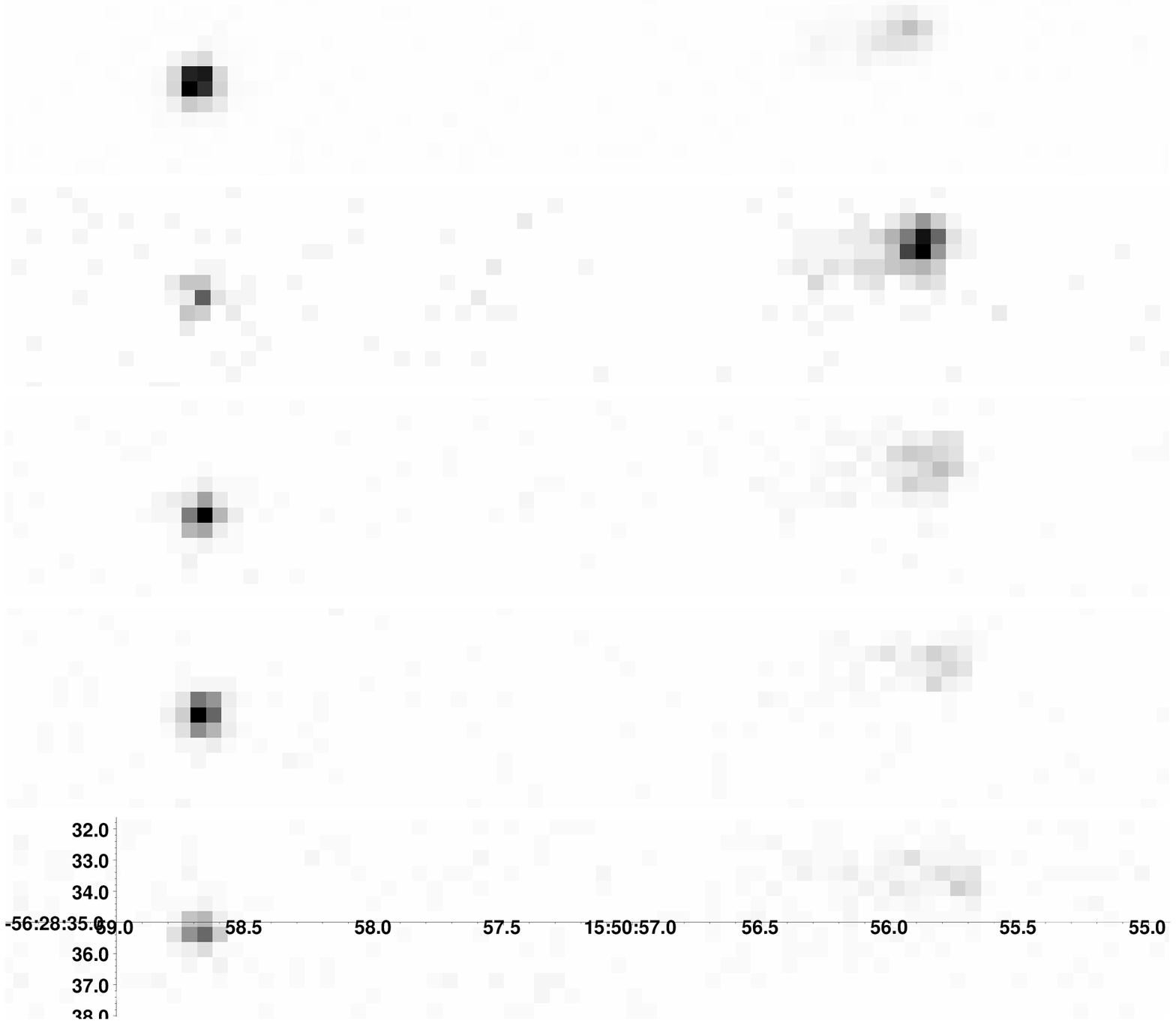}
  \end{minipage}%
  \begin{minipage}[b]{.5\textwidth}
    \centering
    \includegraphics[scale=0.3]{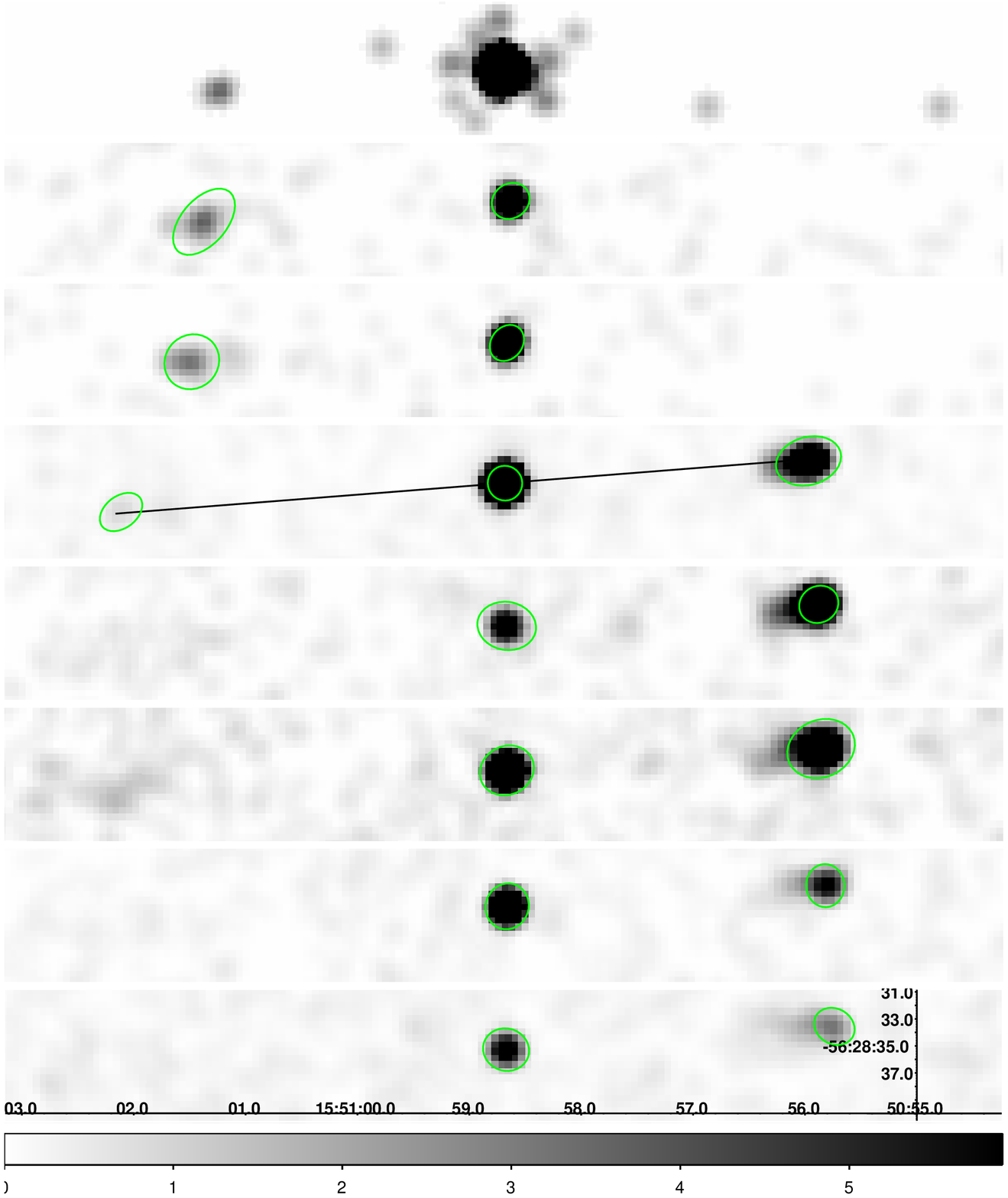}
  \end{minipage}
  \caption{{\it Left}: The Chandra 0.3-8 keV raw image showing XTE J1550-564 and
the western jet. The images are in linear scale and no count
saturation has been set.  \textit{Right}: The smoothed
\textit{Chandra} X-ray images of the eight observations of XTE
J1550-564 and the two jets together. The green elliptical regions
are source emission regions by $wavdect$. Observation 4 shows the
good alignment of the two jets and the central source.}
\end{figure}

From Table 1 and Fig.1, we could see clearly that an X-ray emission source is detected
to the east of the central source in the first four observations and another source is
detected to the west in the last five observations. Calculations also show that these
two sources, when presented in a single combined image, are in good alignment with the
central compact object with an inclination angle of
85.9\textordmasculine$\pm$0.3\textordmasculine. By calculating the average proper
motion, an approximate estimate of deceleration could be seen for both jets.

\section{ENERGY SPECTRUM and FLUX}

Since the emission from the eastern jet has been studied fully (Corbel et al. 2002;
Tomsick et al. 2003), we mainly focus our spectral analysis on the western jet. The
X-ray spectrum in 0.3-8 keV energy band is extracted for each observation of the
western jet. We use a circular source region with a radius of 4\arcsec, an annular
background region with an inner radius of 5$\arcsec$ and an outer radius of 15\arcsec,
for each observation. Instrument response matrices (rmf) and weighted auxiliary
response files (warf) are created using CIAO programs mkacisrmf and mkwarf, and then
added to the spectra. We re-bin the spectra with 10 counts per bin and fit them in
\textit{Xspec}.

The results of spectra fitting with an absorbed power-law model are also shown in table
1. We use the Cash statistic since it is a better method when counts are low. The
absorption column density is fixed to the Galactic value in the direction of XTE
J1550-564 obtained by the radio observations ($N_{H}=9\times10^{21}$cm$^{-2}$) (Dickey
\& Lockman 1990). Our results are quite consistent with previous work by Karret et
al.(2003). The calculated absorbed energy flux in 0.3-8 keV band is comparable to the
value of the eastern jet. The observed flux decayed rather quickly, from
$\sim1.9\times10^{-13}$erg cm$^{-2}$ s$^{-1}$ in March 2002 to only one sixth of this
value in October 2003 (see section 4.2).

\begin{table}[!ht]
\caption{XTE J1550-564 \textit{Chandra} Observations}
\begin{center}
{\small
\begin{tabular}{ccccccccc}
\tableline \noalign{\smallskip}
 &  &  &
\multicolumn{2}{c}{Angular Separations (arcsec)} &
\multicolumn{2}{c}{Powerlaw Fitting for the western jet} \\
\cline{4-5} \cline{6-7}\\
Num & ID   & Date    & Eastern Jet & Western jet    &
Photon Index  & Flux (ergs cm$^{-2}$ s$^{-1}$) \\
\noalign{\smallskip} \tableline \noalign{\smallskip}
1 &679  & 2000 Jun 9  &21.5$\pm$0.5   &     \\
2 &1845 & 2000 Aug 21 &22.8$\pm$0.5   &    &  &  \\
3 &1846 & 2000 Sep 11 &23.4$\pm$0.5   &    &  &  \\
4 &3448 & 2002 Mar 11 &28.6$\pm$0.5   &22.6$\pm$0.5   &1.75$\pm$0.11    &$(1.9\pm0.4)\times10^{-13}$\\
5 &3672 & 2002 Jun 19 &     &23.2$\pm$0.5   &1.71$\pm$0.15    &$(1.6\pm0.3)\times10^{-13}$ \\
6 &3807 & 2002 Sep 24 &29.2$\pm$0.5  &23.4$\pm$0.5  &1.94$\pm$0.17    &$(8.6\pm1.5)\times10^{-14}$\\
7 &4368 & 2003 Jan 28 &     &23.7$\pm$0.5   &1.81$\pm$0.22    &$(5.5\pm1.0)\times10^{-14}$ \\
8 &5190 & 2003 Oct 23 &    &24.5$\pm$0.5   &1.97$\pm$0.20    &$(3.1\pm0.6)\times10^{-14}$\\
\noalign{\smallskip} \tableline

\end{tabular}
}
\end{center}
\end{table}

\section{JET MODEL}

\subsection{Kinematic Model}

In the external shock model for afterglows of GRBs, the kinematic and radiation
evolution could be understood as the interaction between the outburst ejecta and the
surrounding ISM (Rees \& M\'{e}sz\'{a}ros 1992). Microquasar jet systems are also
expected to encounter such interactions. In this section, we describe our attempts
after Wang et al. (2003) in constructing the kinetic and radiation model based on these
models.

We adopt the model of a collimated conical beam with a half opening
angle $\theta_{j}$ expanding into the ambient medium with the number
density $n$. The initial kinetic energy and Lorentz factor of the
outflow material are $E_{0}$ and $\Gamma_{0}$ , respectively. Shocks
should arise as the outflow moves on and heat the ISM, and its
kinetic energy will turn into the internal energy of the medium
gradually. Neglect the radiation loss, the energy conservation
function writes (Huang, Dai, \& Lu 1999):
\begin{equation}\label{a}
(\Gamma-1)M_{0}c^{2}+\sigma(\Gamma_{\textrm{\tiny sh}}^{2}-1)m_{\textrm{\tiny
SW}}c^{2}=E_{0},
\end{equation}
where the first term on the left of the equation represents the kinematic energy of the
ejecta, $\Gamma$ is the Lorentz factor and $M_{0}$ is the mass of the original ejecta.
The second term represent the internal energy of the swept-up ISM, where
$\Gamma_{\textrm{\tiny sh}}$ and $m_{\textrm{\tiny SW}}$ are the corresponding Lorentz
Factor and mass of the shocked ISM respectively, and $ m_{\textrm{\tiny
SW}}=(4/3){\pi}R^{3}m_{\textrm{\tiny p}}n(\theta_{j}^{2}/4)$.

Coefficient $\sigma$ differs from 6/17 to 0.73 for ultra-relativistic and
nonrelativistic jets (Blandford \& McKee 1976). We adopt the approximation of
$\sigma\sim$0.7 after Wang et al.(2003). Equation (1) and the relativistic kinematic
equations
\begin{equation}
(\frac{dR}{dt})_{\textrm{a}}=\frac{\beta(\Gamma)c}{1-\beta(\Gamma)\cos\theta};
(\frac{dR}{dt})_{\textrm{r}}=\frac{\beta(\Gamma)c}{1+\beta(\Gamma)\cos\theta}
\end{equation}
can be solved and give the relation between the projected angular separation $\mu$ and
time $t$. In equations (2), the subscript `a' and `r' represent the approaching and
receding jets in a pair of relativistic jets respectively. $R$ is the distance between
the jet and the source, which can be transformed into the proper motion separation by
$\mu=R\sin\theta/5.3$ kpc, and $\theta$ is the jet inclination angle to the line of
sight. We can get the $\mu-t$ curve numerically with the above equations. To be
consistent with the work previously done to the eastern jet, we choose the same initial
conditions that $\Gamma_{0}=3$, $E_{0}=3.6\times10^{44}$ erg, and
$\theta_{j}=1.\textordmasculine5$. Then the parameters needed to be fit are $n$ and
$\theta$.

In the case of the eastern jet, the number density of the ISM was assumed as a constant
in the whole region outside the central source (Wang et al. 2003). This assumption does
not work well in the case of its western counterpart. The western jet decelerated quite
fast, requiring a local dense environment; however if the ISM is dense everywhere, the
jet will be unable to travel that far from the central point source. As a result, we
consider a model that the ISM density varies as the distance changes. For simplicity,
we test the ideal case that the jet travelled first through a ``cavity" with a constant
velocity and then through a dense region where the jet was decelerated. A new parameter
$r$, the outer radius of the cavity, is introduced and the ISM number density is set to
be a constant $n$ outside this region and zero inside. The fittings improved a lot, but
not well constrained because of the limited number of the data points. A combination of
lightcurve fitting is required to further constrain the model parameters.

\subsection{Radiation Model}
In the standard GRB scenario, the afterglow emission is produced by
the synchrotron radiation or inverse Compton emission of the
accelerated electrons in the shock front of the jets (Wang et al.
2003 and references there). Wang et al.(2003) found that the reverse
shock emission, originating from the electrons of the jet when a
shock moves back through the ejecta, decay rather fast and describe
the data of the eastern jet quite well. We thus take this model in
our work as well.

Assuming the distribution of the electrons obeys a power-law form,
$n{\gamma_{\textrm{\tiny e}}}d\gamma_{\textrm{\tiny
e}}=K\gamma_{\textrm{\tiny e}}^{-p}d\gamma_{\textrm{\tiny e}}$, for
$\gamma_{\textrm{\tiny m}}<\gamma_{e}<\gamma_{\textrm{\tiny M}}$,
the volume emissivity at frequency $\nu'$ in the comoving frame is
given by (Rybicki \& Lightman 1979)
\begin{equation}
j_{\nu'}=\frac{\sqrt{3}q^{3}}{2m_{\textrm{\tiny
e}}c^{2}}(\frac{4{\pi}m_{\textrm{\tiny
e}}c\nu'}{3q})^{\frac{(1-p)}{2}}B_{\pm}^{\frac{(p+1)}{2}}KF_{1}(\nu,\nu'_{\textrm{\tiny
m}},\nu'_{\textrm{\tiny M}}),
\end{equation}
where $F_{1}(\nu,\nu'_{\textrm{\tiny m}},\nu'_{\textrm{\tiny
M}})=\int_{\nu'/\nu'_{\textrm{\tiny M}}}^{\nu'/\nu'_{\textrm{\tiny
m}}}F(x)x^{(p-3)/2}dx$, with $F(x)=x\int_{0}^{+\infty}K_{5/3}(t)$
and $K_{5/3}(t)$ is the Bessel function. The physical quantities in
these equations include $q$ and $m_{\textrm{\tiny e}}$, the charge
and mass of the electron, $B_{\perp}$, the magnetic field strength
perpendicular to the electron velocity, and $\nu'_{\textrm{\tiny
m}}$ and $\nu'_{\textrm{\tiny M}}$, the characteristic frequencies
for electrons with $\gamma_{\textrm{\tiny m}}$ and
$\gamma_{\textrm{\tiny M}}$.

Assuming the reverse shock heats the ejecta at time $t_{0}$ at the radius $R_{0}$, the
physical quantities in the adiabatically expanding ejecta with radius $R$ will evolve
as $\gamma_{\textrm{\tiny m}}=\gamma_{\textrm{\tiny m}}(t_{0})R_{0}/R,
\gamma_{\textrm{\tiny m}}=\gamma_{\textrm{\tiny m}}(t_{0})R_{0}/R$ and
$K=K(t_{0})(R/R_{0})^{-(2+p)}, B_{\perp}=B_{\perp}(t_{0})(R/R_{0})^{-2}$, where the
initial values of these quantities are free parameters to be fitted in the calculation.
With these assumptions, we can then calculate the predicted flux evolution of the jets.
The comoving frequency $\nu'$ relates to our observer frequency $\nu$ by $\nu=D\nu'$,
where $D$ is the Doppler factor and we have $D_{\textrm{\tiny
a}}=1/\Gamma(1-\beta\cos\theta)$ and $D_{\textrm{\tiny r}}=1/\Gamma(1+\beta\cos\theta)$
for the approaching and receding jets respectively. Considering the geometry of the
emission region, the observed X-ray flux in 0.3-8 keV band could be estimated by
\begin{equation}
F(\textrm{0.3-8
keV})=\int_{\nu_{1}}^{\nu_{2}}[\frac{\theta_{j}^{2}}{4}(\frac{R}{d}){\Delta}RD^{3}j_{\nu'}]d\nu,
\end{equation}
where ${\Delta}R$ is the width of the shock region and is assumed to
be ${\Delta}R=R/10$ in the calculation.

To reduce the number of free parameters, we set
$\gamma_{\textrm{\tiny m}}=100$ in our calculation because the
results are quite insensitive to this value. We choose the time that
the reverse shock takes place according to our kinematic model in
section 3.1. Then we fit the data to find out the initial values of
$K$ and $B_{\perp}$.

Next step, we combine the kinematic and radiation fitting together.
We know that the energy and the number density of the gas in the
pre-shock and post-shock regions are connected by the jump
conditions $n'=\zeta(\Gamma)n$ and $e'=\eta(\Gamma)nm_{\textrm{\tiny
p}}c^{2}$, where $\zeta(\Gamma)$ and $\eta(\Gamma)$ are coefficients
related to the jet velocity. Therefore if we assume the shocked
electrons and the magnetic field acquire constant fractions
($\epsilon_{\textrm{\tiny e}}$ and $\epsilon_{\textrm{\tiny B}}$) of
the total shock energy, we have $\gamma_{\textrm{\tiny
m}}=\epsilon_{\textrm{\tiny e}}(p-2){m_{\textrm{\tiny
p}}}(\Gamma-1)/(p-1){m_{\textrm{\tiny e}}}$,
$K=(p-1)n'\gamma_{\textrm{\tiny m}}^{p-1}$, and
$B_{\perp}=\sqrt{8\pi\epsilon_{\textrm{\tiny B}}e'}$.

If we further assume that the $\epsilon_{\textrm{\tiny e}}$ of the
eastern and the western jets is the same, we may infer that
$K\propto{e'}\propto{n}$ for the two jets. As a result, we search
for the combination of parameters that could satisfy the kinematic
and radiation fitting, as well as the relationship $K_{\textrm{\tiny
e}}/K_{\textrm{\tiny w}}{\sim}n_{\textrm{\tiny e}}/n_{\textrm{\tiny
w}}$.

A set of parameters has finally been found (Please refer to the \textit{Left} panel in
Fig.2). The boundary of the cavity lies at $r\sim$14$\arcsec$ to the east and
$\sim$18$\arcsec$ to the west of the central source. The corresponding number density
of the ISM outside this boundary is $\sim$0.00675 cm$^{-3}$ and $\sim$0.21 cm$^{-3}$,
respectively. Both values are lower than the canonical ISM value of $\sim$1 cm$^{-3}$,
although the value in the western region is much higher than that in the eastern
region. The electron energy fraction relationship is satisfied as $K_{\textrm{\tiny
e}}/K_{\textrm{\tiny w}}{\sim}n_{\textrm{\tiny e}}/n_{\textrm{\tiny w}}\sim0.03$. On
the other hand, the other relation concerning the magnetic field strength could not be
satisfied simultaneously by these parameters. Although the cavity radius and the number
density are allowed to vary significantly, the best fitted magnetic field strength
remains quite stable($\sim$0.4-0.6 mG). One possible interpretation for this is that
the equipartition parameter varies as the physical conditions of the jet varies; an
alternative explanation may involve the {\it in situ} generation (or amplification) of
the magnetic field.

\begin{figure}
   \begin{minipage}[b]{.5\textwidth}
    \centering
    \includegraphics[scale=0.4]{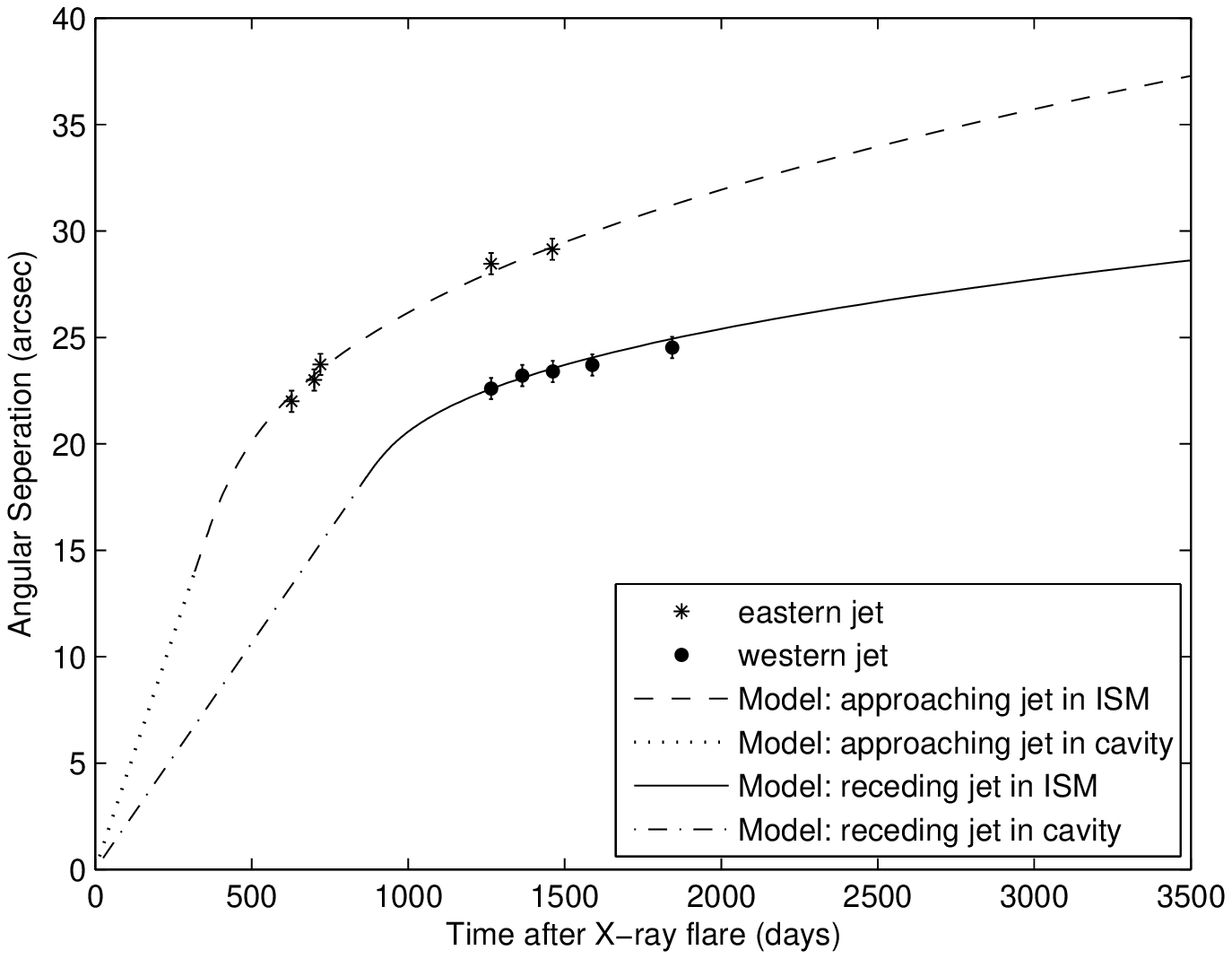}
  \end{minipage}
  \begin{minipage}[b]{.5\textwidth}
    \centering
    \includegraphics[scale=0.4]{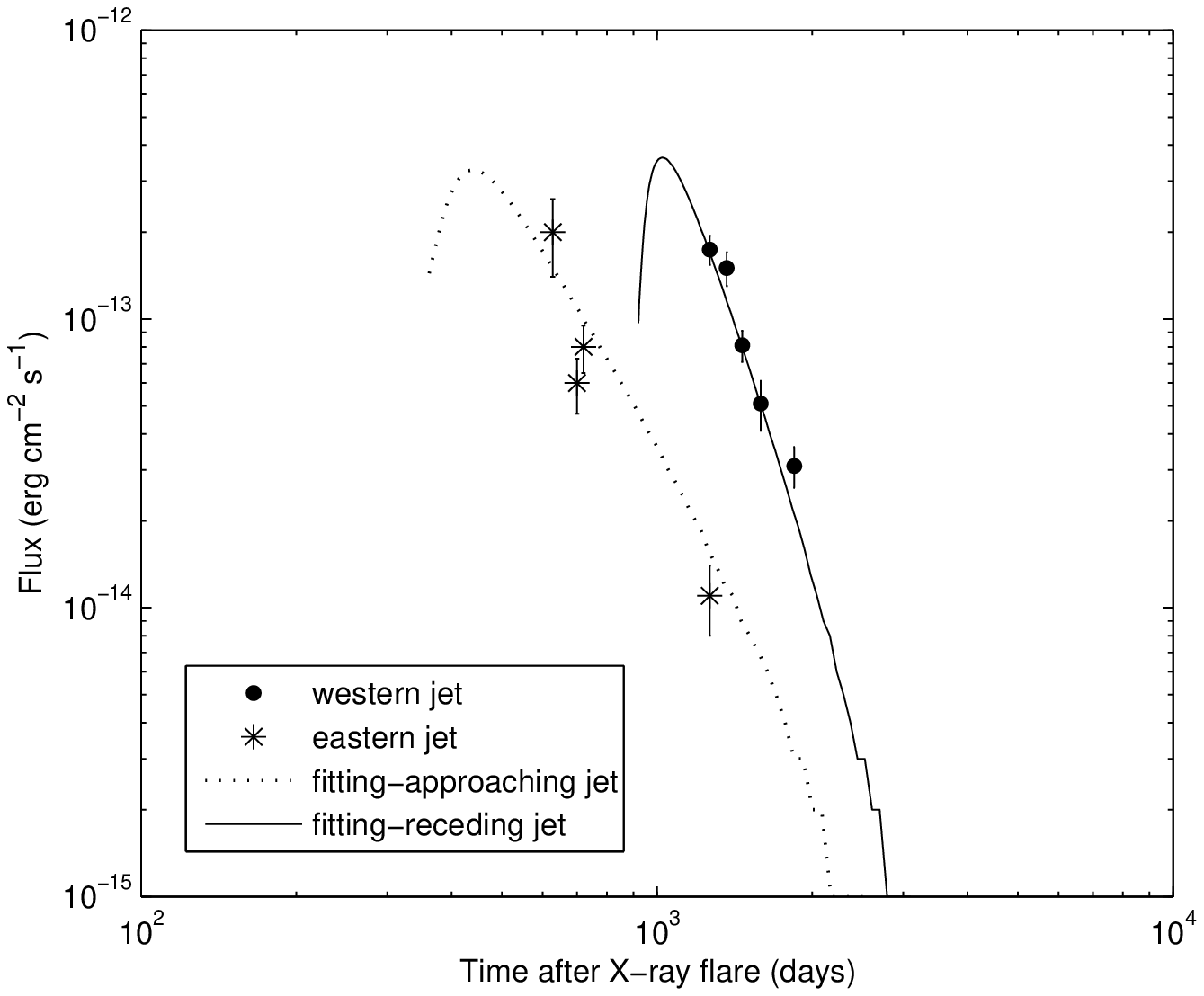}
  \end{minipage}
  \caption{{\it Left}: Proper motion fitting with asymmetric
cavity. Parameters: $\theta$=68\textordmasculine, $r_{\textrm{\tiny
e}}$=14 arcsec, $r_{\textrm{\tiny w}}$=18 arcsec, $n_{\textrm{\tiny
e}}$=0.00675 cm$^{-3}$, $n_{\textrm{\tiny w}}$=0.21 cm$^{-3}$. {\it
Right}: Reverse shock emission fitting to the X-ray light curve of
the jets. $K_{\textrm{\tiny e}}$=0.09 cm$^{-3}$, $K_{\textrm{\tiny
w}}$=0.3 cm$^{-3}$, $B_{\textrm{\tiny e}}$=0.5 mG, $B_{\textrm{\tiny
w}}$=0.4 mG.}
\end{figure}

\section{Conclusion and Discussions}
A GRB external shock model shows that a large scale cavity exists outside XTE
J1550-564. This model has also been applied to another X-ray transient H 1743-322.
Chandra X-ray and ATCA radio observations of H 1743-322 from 2003 November to 2004 June
revealed the presence of large-scale ($\sim$0.3 pc) jets with velocity $v/c\sim0.8$
(Rupen et al. 2004; Corbel et al. 2005). Deceleration is also confirmed in this system.
The external shock model describes the data of this source well. A cavity of size
$\sim$0.12 pc most likely exists, but the conclusion is not firm in this case. Even if
there is no vacuum cavity, the ISM density is found to be very low($\sim3\times10^{-4}$
cm$^{-3}$), compared to the canonical Galactic value.

These studies led us to the suggestion that in microquasars the interactions between
the ejecta and the environmental gas play major roles in the jet evolution and the low
density of the environment is a necessary requirement for the jet to develop to a long
distance. We find that microquasar jets can be classified into roughly three groups:
small scale moving jets, large scale moving jets and large scale jet relics. For the
first type, the ``small jets", only radio emissions are detected. The jets are always
relatively close to the central source and dissipate very quickly, including GRS
1915+105 (Rodr\'{\i}guez \& Mirabel 1999; Miller-Jones et al. 2007), GRO J1655-40
(Hjellming \& Rupen 1995), and Cyg X-3 (Marti et al. 2001). The typical spatial scale
is 0$\sim$0.05 pc and the time scale is several tenths of days. No obvious deceleration
is observed before the jets become too faint. For the second type, the ``large jets",
both X-ray and radio detections are obtained, at a place far from the central source
several years after the outburst. Examples are XTE J1550-564, H1743-322, and GX 339-4
(Gallo et al.2004). The typical jet travelling distance for this type is 0.2$\sim$0.5
pc from the central engine and deceleration is clearly observed. The last type, the
``large relics", is a kind of diffuse structures observed in radio, optical and X-ray
band, often ring or nebula shaped that are not moving at all. In this class, some well
studied sources, Cygnus X-1 (Gallo et al.2005), SS433 (Dubner el al.1998), Circinus X-1
(Stewart et al. 1993), 1E~1740.7-2942 (Mirabel et al. 1992) and GRS 1758-258
(Rodr\'{\i}guez et al. 1992) are included. The typical scale for this kind is 1$\sim$30
pc, an order of magnitude larger than the second type. The estimated lifetime often
exceeds one million years, indicating that they are related to previous outbursts.

From these properties, it is reasonable to further suggest a consistent picture
involving all the sources together. We make a conjecture that large scale cavities
exist in all microquasar systems. The ``small jets" observed right after the ejection
are just travelling through these cavities. Since there are few or none interactions
between the jets and the surrounding gas in this region, the jets travel without
obvious deceleration. The emission mechanism is synchrotron radiation by particles
accelerated in the initial outburst. The emissions of jets decay very quickly and are
not detectable after several tenths of days. In some cases (e.g. XTE J1550-564), the
cavity has a dense (compared to the cavity interior) boundary at some radius and the
interactions between the jets and the boundary gas heat the particles again and thus
make the jets detectable again. Those are the ``large jets". The emission mechanism
then is synchrotron radiation by the re-heated particles in the external shocks. Then,
after these interactions, the jets lost most of their kinetic energy into the ISM
gradually, causing the latter to expand to large scale structures, the ``large relics",
in a comparatively long time (several millions of years).

The creation of the cavities is not clear at this stage. Possible mechanism may involve
supernovae explosions, companion star winds or disk winds. Since some of the sources
most likely have never had supernovae before and the winds from the companion stars are
not strong enough, the accretion disk winds may be the most plausible possibility.
However, these assumptions all require further observations to justify.

Microquasars are powerful probes of both the central engine and their surrounding
environment. More studies of their jet behaviors may give us information on the ISM gas
properties, as well as the ejecta components. It will provide insights of the jet
formation process and offer another approach into black hole physics and accretion flow
dynamics.


\acknowledgements We thank Dr. Yuan Liu, Shichao Tang and Weike Xiao
for useful discussions and Xiangyu Wang for providing the model
codes. SNZ thanks the SOC and LOC for great effort in organizing
this conference. This study is supported in part by the Ministry of
Education of China, Directional Research Project of the Chinese
Academy of Sciences under project No. KJCX2-YW-T03 and by the
National Natural Science Foundation of China under project No.
10521001, 10733010 and 10725313.



\end{document}